\documentclass[twocolumn]{aa}			
\usepackage{graphicx}

\usepackage{txfonts}
\usepackage{natbib}
\bibpunct{(}{)}{;}{a}{}{,}
\def\gapprox{{_>\atop{^\sim}}}
\def\lapprox{{_<\atop{^\sim}}}

\def\cmmd{\rm {cm^{-3}}}
\def\cmmt{\rm {cm^{-2}}}

\def\s-1{\rm {s^{-1}}}
\def\twco{$^{12}$CO}

\def\hop{H$_3$O$^+$}
\def\hpo{H$_3$O$^+$}
\def\hto{H$_2$O}
\def\hcop{HCO$^+$}
\def\hc3n{HC$_3$N}

\def\kms {\hbox{${\rm km\,s}^{-1}$}}

\usepackage{url}

\begin{document}
   \title{H$_3$O$^+$ Line Emission from Starbursts and AGNs}
 
   \author{S. Aalto\inst{1}
        \and
        F. Costagliola\inst{1}
	\and
	F. van der Tak\inst{2,3}
	\and
	R. Meijerink\inst{4}
          }

   \offprints{S. Aalto}

   \institute{Department of Earth and Space Sciences, Chalmers University of Technology, Onsala Observatory,
              SE-439 92 Onsala, Sweden\\
              \email{saalto@chalmers.se}
        \and
		SRON, Netherlands Institute for Space Research, Landleven 12, NL-9747 AD Groningen, The Netherlands
	 \and
		Kapteyn Astronomical Institute, University of Groningen, The Netherlands
	\and
		Leiden Observatory, Leiden University, P.O. Box 9513, NL-2300 RA, Leiden, The Netherlands
             }

   \date{Received x; accepted y}


    \abstract
  {The \hop\ molecule probes the chemistry and the ionization rate of dense circumnuclear gas in galaxies.}
   {We use the \hop\ molecule to investigate the impact of starburst and AGN activity on the chemistry of the molecular interstellar medium.}
   {Using the JCMT, we have observed the 3$^{+}_{2} - 2^{-}_{2}$ 364~GHz line of p-\hop\ towards the centers of seven active galaxies.}
   {We have detected p-\hop\ towards IC~342, NGC~253, NGC~1068, NGC~4418, and NGC~6240. Upper limits were
obtained for IRAS~15250 and Arp~299. We find large \hop\ abundances ($N$(\hop)/$N$(H$_2$)$\gapprox 10^{-8}$) in
all detected galaxies apart from in IC~342 where it is about one order of magnitude lower. We note, however, that 
uncertainties in $N$(\hop) may be significant due to lack of definite information on source size and excitation. We furthermore
compare the derived $N$(\hop) with $N$(\hcop) and find that the \hop\ to \hcop\ column density ratio is large in
NGC~1068 (24), moderate in NGC~4418 and NGC~253 (4-5), slightly less than unity in NGC~6240 (0.7) and lowest in IC~342 (0.2-0.6).
We compare our results with models of X-ray and photon dominated regions (XDRs and PDRs).}
{For IC~342 we find that a starburst PDR chemistry can explain the observed H$_3$O$^+$ abundance. For the other galaxies,
the large H$_3$O$^+$ columns are generally consistent with XDR models. In particular for NGC~1068 the elevated $N$(\hop)/$N$(\hcop) ratio suggests
a low column density XDR. For NGC~4418 however, large \hc3n\ abundances are inconsistent with the XDR interpretation.
An alternative possibility is that H$_3$O$^+$ forms through \hto\ evaporating off dust
grains and reacting with \hcop\ in warm, dense gas. This scenario could also potentially fit the results for NGC~253. Further studies of
the excitation and distribution of \hop\ - as well as Herschel observations of water abundances - will help to further constrain the models.
}
\titlerunning{Extragalactic H$_3$O$^+$}
\maketitle

%

\section{Introduction} 

Molecular line emission is an important tool for probing the highly obscured
inner regions of starburst galaxies and buried AGNs.
Line ratios within and between species help determine physical conditions 
and chemistry of the gas, which provide essential clues to the type and evolutionary stage of the
nuclear activity. Important extragalactic probes of cloud
properties include molecules such as HCN, HCO$^+$, HNC, CN, and HC$_3$N
\citep[e.g.][]{costagliola10,krips08,gracia08,loenen08,imanishi04,gao04,aalto02} that trace the dense ($n \gapprox 10^4$ $\cmmd$)
star forming phase of the molecular gas. HCN, HNC, HCO$^+$ and CN are all species that
can be associated both with photon dominated regions (PDRs) \citep[e.g.][]{tielens85} in starbursts and
X-ray dominated regions (XDRs) \citep[e.g.][]{maloney96,meijerink05} surrounding active galactic nuclei (AGN).
In contrast, \hc3n\ requires shielded dense gas to survive in significant abundance since it is destroyed
by UV and particle radiation (e.g. reactions with the ions C$^+$ and He$^+$)\citep[e.g.]{prasad80,rodriguez-franco98}.
These ions are expected to be abundant in for example XDRs.
Thus HC$_3$N line emission may identify galaxies where the starburst is in the early, embedded, stage of its evolution.
There is still, however, substantial dichotomy in the interpretation of the line emission from the above molecules and therefore
new tracer species to combine with existing information are important.

The hydronium ion \hop\ is a key species in the oxygen chemistry of dense molecular clouds, and
useful as a measure of the ionization degree of the gas \citep{phillips92}. Since \hop\ formation requires \hto\ to exist
in the gas-phase, the \hop\ molecule acts as a natural filter to select hot ($T>$ 100 K) molecular gas,
and therefore traces more specific regions than molecules usually surveyed towards other galaxies.
The chemistry of this filtering is the evaporation of icy grain mantles at $T \approx$ 100 K \citep{vandertak06b},
or that \hto\ forms in the gas phase at high ($T> 300$ K) temperatures.

Observations of \hop\ emission will therefore help probe the location of dense, warm and active gas in galactic nuclei.
Combined with information on the \hto\ abundance, these observations may further be
used to trace the ionization rate by cosmic rays (produced in supernovae,
i.e., starbursts) and/or X-rays (from an AGN). Studies by \citet{vandertak06a} have demonstrated this use
of \hop\ for Sgr~B2 in the Galactic Centre. Furthermore, recent Herschel observations of Galactic sources
show 984~GHz 0$^{-}_{0} - 1^{+}_{0}$ \hop\ absorption in the diffuse gas towards
G10.6-0.4 (W31C) \citep{gerin10} and 1.03 - 1.63 THz \hop\ emission from the high mass star forming region
W3IRS5 \citep{benz10}. In general \hop\ abundances agree well with expectations from PDR models and - together with H$_2$O$^+$ and
OH$^+$ - provide important further insight into gas phase oxygen chemistry.

Recently we have detected 3$^{+}_{2} - 2^{-}_{2}$ \hop\ in the nearby starburst M~82 and the ultraluminous galaxy Arp~220 \citep{vandertak08}.
Derived column densities, abundances, and \hop/\hto\ ratios indicate ionization rates similar to or even exceeding
that in the Galactic Center. In M~82 the extended evolved starburst (PDR) is a likely source of this ionization rate while,
for the ULIRG Arp~220, an AGN-origin (XDR) is suggested.

In the XDR and PDR models grain-processing is not taken into account since the chemistry is taking place in the gas phase.
However, a formation route for \hop\ involving the evaporation (or removal by shocks) of \hto\ from grain surfaces need also
to be considered. For example, evaporating \hto\ reacting with \hcop\ may provide an important source of \hop\
\citep[e.g.][]{phillips92,vandertak00}.

We have used the James Clerk Maxwell Telescope (JCMT) in Hawaii to observe the $J_{\rm K} =3^{+}_{2} - 2^{-}_{2}$ 364 \hop\ line 
(upper level energy $E_{\rm u}$= 139 K) in seven 
starburst and active galaxies which cover a range of environments. Our goal is to use \hop\ as a tracer of gas properties
in galactic nuclei and to see if \hop\ can serve as a diagnostic tool to
distinguish AGN from starburst activity.
In sections \ref{s:sample}, \ref{s:obs} and \ref{s:res} the sample, observations and their results are presented.
\hop\ line parameters are presented in section~\ref{s:line} and in section~\ref{s:column} \hop\ column densities and
fractional abundances are calculated. In sections~\ref{s:xdr} and~\ref{s:hcop} \hop\ abundances in the context of X-ray and UV
irradiated models are discussed and in \ref{s:grain} the potential importance of grain chemistry. In the last
section, \ref{s:future}, we present a brief future outlook.


\section{The sample}
\label{s:sample}

\begin{table}
\caption{\label{t:sample} Sample galaxies$^a$.}
\begin{tabular}{lllcc}
 & \\
\hline
\hline \\

{\bf Galaxy} & $\alpha$\,(J2000) & $\delta$\,(J2000) & $L_{\rm FIR}$ & $D$\\

& hh~mm~ss & $^{\circ}$~$'$~$''$ & L$_{\odot}$ & Mpc \\

\hline 
&\\

{\bf IC~342} & 03:46:48.00  & +68:05:46.0 & $6 \times 10^8$ L$_{\odot}$ & 1.8 \\

\\
{\bf NGC~253} & 00:47:33.12 & -25:17:17.59 & $2.1 \times 10^{10}$ & 2.6 \\

\\
{\bf NGC~1068}$^d$ &  02:42:40.71  &  -00:00:47.8 & $2 \times 10^{11}$  & 14.4 \\ 

\\
{\bf NGC~4418} & 12:26:54.63   &    -00:52:39.6 & $8 \times 10^{10}$  & 27.3 \\

\\
{\bf NGC~6240} &  16:52:58.89   &    +02:24:03.4  & $3.5\times 10^{11}$ & 107 \\

\\

{\bf IRAS~15250} &  15:26:59.40   &    +35:58:38.0 & $1.12 \times 10^{12}$  & 244 \\

\\

{\bf Arp~299} &  11:28:33.13  &  +58:33:58.0 & $8 \times 10^{11}$ & 42 \\
&\\
\hline 
\\
\end{tabular}

a) References: IC~342: \citet{downes92}; NGC~253: \citet{strickland04}; NGC~1068: \citet{telesco84};
NGC~4418: \citet{ridgway94}; NGC~6240: \citet{yun02}; IRAS15250: \citet{veilleux99}; Arp~299: \citet{aalto97}

\end{table}

We have selected a sample consisting of seven nearby luminous starburst and AGNs - and
one distant ULIRG. The galaxies are selected from their bright HCN line emission.
From our previous experience with extragalactic \hop\ we knew that the line is weaker
than the standard high density gas tracers such as HCN and \hcop\ so we restricted ourselves to a relatively small
sample of seven objects (coordinates, FIR luminosities and distances are presented in Table~\ref{t:sample}):\\

{\it IC~342} is a nearby barred Scd galaxy of moderate luminosity (central 400~pc has
$L_{\rm FIR}$ of $6 \times 10^8$ L$_{\odot}$) with a central starburst. Within its central 300~pc (30\arcsec)
two molecular arms end in a clumpy central ring of dense gas \citep[e.g.][]{downes92} which surrounds a young
star cluster. The ring is suggested to outline the $X2$ orbits in a larger-scale bar. The chemistry of IC~342 has been
investigated in detail in a high-resolution study by \citet{meier05}. They find that the chemistry
in the ring is a mixture of PDR-dominated regions and regions of younger star-forming clouds. The chemistry in the
bar/arms is dominated by shocks as shown by CH$_3$OH \citep{meier05} and SiO \citep{usero06}.
Five (A-E) giant molecular clouds (GMCs) are found in the ring and arms. The 13\arcsec\ JCMT beam of our \hop\
observations is pointed towards the region of GMC~B - but also includes GMCs A and E. {\it GMCs B and C are the sites
of young ( a few Myr) star formation and are also the regions where incoming molecular clouds meet the ring. GMC A has more
PDR characteristics. }
The dust temperature of IC~342 is estimated to 42~K \citep[e.g.][]{becklin80}.\\

{\it NGC~253} is also a nearby barred galaxy located in the Sculptor group with a compact nuclear starburst and a
IR luminosity that appears to originate in regions of intense massive star formation within its central few hundred
parsecs \citep{strickland04}. 
From their 2~mm spectral scan \citet{martin06} suggest that the chemistry of NGC~253 shows strong similarities
to that of the Galactic Center molecular region, which is thought to be dominated by low-velocity shocks.
High resolution SiO observations show bright emission resulting from large scale shocks as well as gas entrained in a
nuclear outflow \citep{garcia-burillo00}.
Note also that it is suggested that NGC~253 is a galaxy in which a strong starburst and a weak AGN coexist
\citep[e.g.][]{weaver02,muller-sanchez10}.
High resolution observations of HCN and \hcop\ 1--0 \citep{knudsen07} show strongly centrally concentrated emission.
The 13\arcsec\ JCMT beam covers the bulk of the nuclear emission from HCN and \hcop. The central dust temperature
of NGC~253 is estimated to 50~K \citep{melo02}.\\

{\it NGC~1068} is the nearest example of a type~2 Seyfert galaxy luminous in the infrared. Surrounding the AGN there is a 4\arcsec\ circumnuclear molecular ring or -disk (CND) and on a larger scale there is
a NIR stellar bar 2.3~kpc long. This bar is connected to a large-scale, molecular starburst ring that contributes
about half the bolometric luminosity of the galaxy \citep[e.g.][]{telesco84, scoville88, tacconi94, helfer95, tacconi97}. 
Bright HCN 1--0 line emission is observed towards the CND  while the \hcop\ 1--0 emission
is relatively fainter by a factor of 1.5 \citep[e.g.][]{kohno01}. 
Only the CND and a fraction of the bar is covered by our JCMT beam and we adopt a source size of 2\arcsec\ for the CND.
The dust of the inner 4\arcsec\ appears to show a strong temperature gradient - from about 800~K in the very inner region
to 150 - 275~K at a distance of 0.\arcsec 8 or greater \citep{tomono06}. \citet{alloin00} find temperatures of about 150~K
200~pc from the nucleus. There is a also radio jet from the nucleus which falls into our JCMT beam \citep[e.g.][]{wilson83}.\\

{\it NGC~4418} is a luminous, edge-on, Sa-type galaxy with a deeply dust-enshrouded nucleus \citep{spoon01}.
NGC~4418 is a FIR-excess galaxy with a logarithmic IR-to-radio continuum ratio ($q$) of 3 \citep{roussel03}. This excess may be caused by either a young pre-supernova starburst or a buried AGN \citep{aalto07b, roussel03, imanishi04}. Unusually luminous HC$_3$N line emission \citep{aalto02,aalto07b,costagliola10} has been interpreted as a signature of young starburst activity.
Mid-IR intensities are indicative of dust temperatures of 85 K \citep{evans03} inside a radius of 50 pc (0.\arcsec 5). The IR luminosity-to-molecular gas mass ratio is high for a non-ULIRG galaxy indicating that intense, compact activity is hidden behind the dust. Recent high resolution
imaging of CO 2--1 (Costagliola et al. in prep.) indicate a molecular source size of 0.\arcsec 5.\\

{\it NGC~6240} is an infrared luminous merger of two massive spiral galaxies. Two AGN/LINER nuclei are separated by 1\arcsec\ and the
bulk of the molecular gas has piled up between the two nuclei \citep{iono07,tacconi99}. Luminous \hcop\ 4--3 emission
is also emerging from in between the two nuclei where the H$_2$ emission is also located. The mid-IR sources (24.5 $\mu$m)
are associated with the two X-ray nuclei - the brightest by far being the southern nucleus ($T$=55-60 K)\citep{egami06}.
It has been suggested that the medium between the two nuclei is dominated by starburst superwinds from the southern nucleus
\citep{ohyama00} and the molecular gas is indeed highly turbulent \citep{iono07,tacconi99}.
The JCMT beam encompasses the entire molecular structure of this galaxy and we adopt a source size of 1\arcsec\ for the \hop\ emission.\\

{\it IRAS~15250+3609} is a relatively distant ultraluminous galaxy - probably a major merger with a dominant bright nucleus.
A tidal feature appears to emerge from the southwest side of the nucleus and turns around on the eastern
side to create an enormous closed ring, 27~kpc in diameter. The optical spectrum is a composite of H~II and LINER
features \citep{veilleux99}.  IRAS~15250 exhibits deep silicate absorption features \citep{spoon06} indicating a deeply
enshrouded nucleus. Our JCMT beam covers the whole optical galaxy.\\

{\it Arp~299} is an IR-luminous merging system of two galaxies, IC~694 and NGC~3690. Strong \twco\ emission has been detected from the nuclei of IC~694 and NGC~3690 and from the interface between the two galaxies \citep[e.g.][]{sargent91, aalto97}. The two nuclei, as well as the western overlap region, currently undergo intense star formation activity
\citep[e.g.][]{gehrz83}. Bright HCN 1--0 emission is emerging from both nuclei as well as from the
overlap region \citep[e.g.][]{aalto97}. Water emission from NGC~3690 suggests that it also harbours an AGN \citep{tarchi07}.
To cover all three regions a map was carried out.


\section{Observations}
\label{s:obs}

\begin{figure}
\center
\resizebox{5.5cm}{!}{\includegraphics[angle=0]{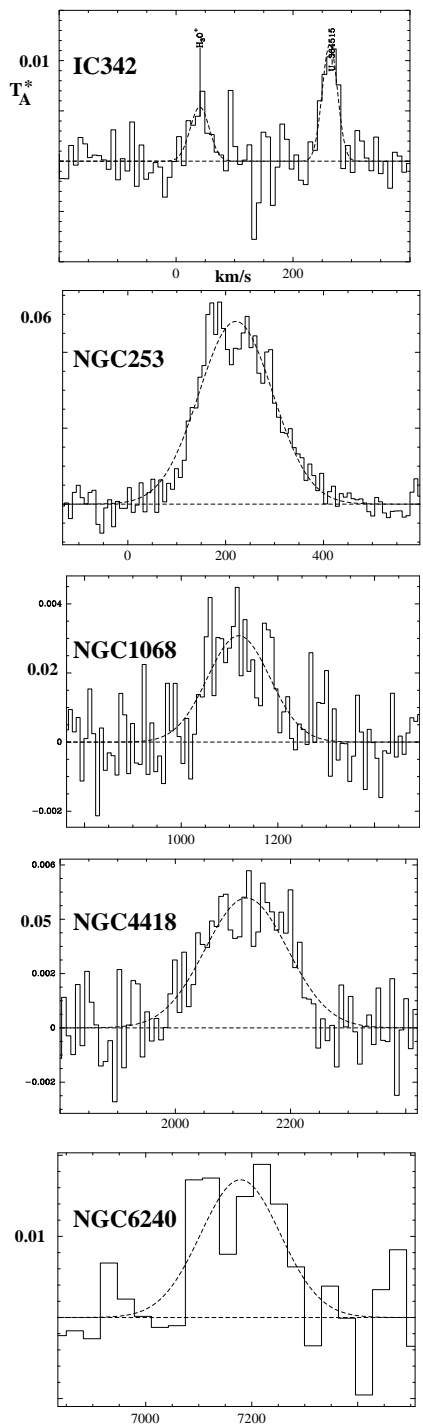}}
\caption{\label{f:spec} Observed spectra of \hop. The intensity scale is in T$_A^*$, not corrected for the JCMT
beam efficiency, that at this frequencies is 0.7. The parameters of the detected lines were obtained by Gaussian fits,
shown on the plots as dashed lines. }
\end{figure}

The 364.7974~GHz $J_{\rm K}=3^{+}_{2} - 2^{-}_{2}$ line of \hpo\ was observed towards the sample galaxies in February and August 2008, with the
James Clerk Maxwell Telescope (JCMT\footnote{The JCMT is operated by the Joint Astronomy Center on behalf of
the Science and Technology Facilities Council of the United Kingdom,
the Netherlands Organization for Scientific Research, and the National
Research Council of Canada.}) on Mauna Kea, Hawaii.
We used the 16 elements heterodyne array HARP. Each of the receptors of the array has a beam size of 14\arcsec at
345 GHz and the sources were placed in the receptor labeled H05\footnote{For more information see: \url{www.jach.hawaii.edu/JCMT/spectral_line/Instrument_homes/HARP/HARP.html}}. 
The back end was the Auto-Correlation Spectrometer and Imaging System (ACSIS), providing 1.0~GHz
bandwidth in 2048 channels and a maximum resolution of 0.4~\kms.
 In order to optimize baseline stability, double
beam switching at a rate of 1~Hz was used with an offset of 120\arcsec.
Weather conditions were
optimal, with about 1~mm of precipitable water vapor, leading to a typical noise rms of 2 mK at 10~km/s resolution.
Telescope pointing was checked every hour on the CO line emission
of nearby AGB stars and always found to be within 2\arcsec.
Data were extracted into fits files with the starlink software and analyzed in CLASS. Linear baselines were
subtracted, line parameters were calculated by fitting
Gaussian profiles to the spectra. The resulting spectra are shown
in Fig.~\ref{f:spec}. Throughout
this paper, velocities are in the heliocentric frame and redshift is
computed using the radio convention.

\section{Results}
\label{s:res}

\subsection{Line intensities and line widths}
\label{s:line}

Integrated line intensities, line widths and fitted velocities (heliocentric) for the
seven observed galaxies can be found in Table~\ref{t:line}. p-\hpo\ line emission was detected 
towards IC~342, NGC~4418, NGC~253, NGC~1068 and NGC~6240, while Arp~299 and IRAS~15250 were not detected.

In general it is found that line widths and shapes (Fig.~\ref{f:spec}) agree well with those found for HCN and \hcop\
in the central regions of the systems \citep{nguyen92,aalto07b,knudsen07}. 
The most noteworthy deviation from this is NGC~6240 where
the linewidth of HCN and \hcop\ \citep{greve09} is more than a factor of two greater than what we find for \hop.
A possible explanation for this could be the weak \hop\ signal resulting in a moderate
signal-to-noise profile. For NGC~1068 the line width agrees well with that of \hcop\ 4--3 but
is somewhat narrower than what is observed for HCN 4--3 (202 \kms) \citep{jp09}.

For IC~342 an unidentified (U) line was detected in February 2008 - but it did not appear again in August the same year. 
The line is somewhat narrower than the \hop\ line and appear at a velocity of 263 \kms\ (rest frequency 364.477 GHz).
It is possible that it is an atmospheric O$_3$ line.

\begin{table}
\caption{\label{t:line} \hop\ Line Results$^a$.}
\begin{tabular}{lcccc}
 & \\
\hline
\hline \\

{\bf Galaxy} &  $\int T_{\rm A}^*$ & $V_{\rm line}$  & $\delta V$  & $T_{\rm A}^*$ \\

& [K km/s] & [km/s] & [km/s] & [mK] \\

\hline 
&\\
{\bf IC~342} &\\
{\it 08-2008:}$^b$ & \\
\hop &  $0.30 \pm 0.07$  & $39 \pm 10$ &  $75 \pm 18$ &  3.7\\
{\it 02-2008:}$^b$ & \\
\hop &  $0.22 \pm 0.06$ &  $40 \pm 5$ &  $39 \pm 13$ &  5.4\\
(U-line & $0.38 \pm 0.05$ & $263 \pm 2$ &   $30\pm 5$ & 11.6) \\

\\
{\bf NGC~253}  & $9.35 \pm 0.18$ &   $221 \pm 2$ &   $180 \pm 4$ &  48.8 \\

\\
{\bf NGC~1068}$^d$ & $0.52 \pm 0.05$ &  $1120\pm 8$ & $160\pm 20$ &  3.0\\ 
\\
{\bf NGC~4418}  & $1.02\pm  0.06$ &  $2129\pm 5$ &  $171\pm 11$ &  5.6 \\

\\
{\bf NGC~6240}   & $0.32 \pm 0.05$ &  $7179\pm 15$ &  $177 \pm 27$ &  1.7 \\

\\

{\bf IRAS~15250}  & $<$ 0.012$^c$ & $\dots$ & $\dots$ & $<$1.0 \\

\\

{\bf Arp~299} & $<$ 0.020$^d$ & $\dots$ & $\dots$ & $<$2.0 \\
&\\
\hline 
\\
\end{tabular}

a) Integrated line intensities and peak line intensities are given in $T_{\rm A}^*$, $\eta_{rm mb}$ is 0.65-0.7
and we have adopted 0.7 here for conservative values of $N$(\hop).
b) Data reduction of \hop\ and U-line taken at JCMT 02-2008 and 08-2008. 
c) 1$\sigma$ upper limit for a FWHM linewidth of 500 km/s.
d) 1$\sigma$ upper limit for a FWHM linewidth of 300 km/s. The pointing center of HARP was in between the
two nuclei. We added the spectra in the three emission regions of Arp~299 covered by HARP to produce a spectrum
with a 1$\sigma$ upper limit of 2~mK.

\end{table}

\subsection{Column densities, excitation and abundances}
\label{s:column}

Since we have no direct information on the excitation of the \hop\ molecule we make some assumptions
that will have to be tested in future observations. The critical density of the \hpo\ transition is 
high (about $n_{\rm crit} = 10^6$ $\cmmd$ \citep{phillips92}) and the low reduced mass of the
\hop\ molecule makes its excitation very sensitive to radiative pumping by dust.
We therefore assume that the coupling between the colour temperature of the IR emission and the excitation temperature
of \hop\ holds as long as the dust colour temperature is above 30~K (see discussion in \citet{vandertak08}).

For each galaxy we ran a RADEX model \citep{vandertak07} with temperatures ranging from $T$=10-500 K.
In these calculations, the volume density is set to an arbitrary large value, so that the excitation of \hop\ is thermalized.
The adopted excitation temperature is essentially the radiation temperature and a radiative excitation with one
excitation temperature is simulated. We therefore do not use the full non-LTE capacity of RADEX.

\begin{figure*}
\resizebox{17cm}{!}{\includegraphics[angle=0]{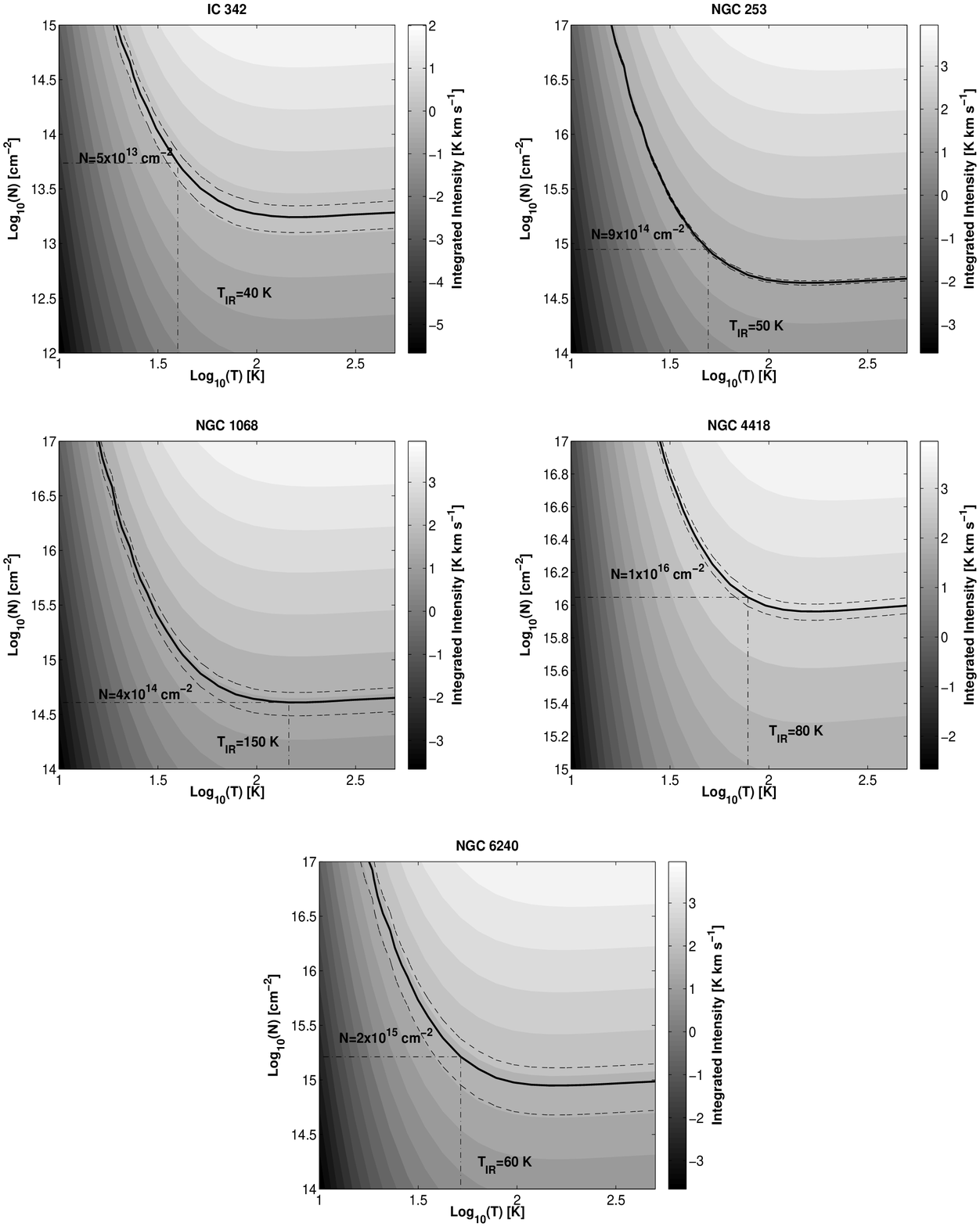}}
\caption{\label{f:radex} On the x axis we find the temperature of the
exciting background radiation and on the y axis the column density of \hpo. For each galaxy, the brightness temperature
of the \hpo\ transition is shown in a gray scale map. The thick black contour marks the observed value, with one~sigma
errors are marked by dashed lines. Background temperatures estimated by IR observations are also reported and used to
derive a column density consistent with the observed brightness temperature. Note that the column density fits include
only the para column.}
\end{figure*}

The results are shown in Fig.~\ref{f:radex}. Each plot has temperature (kinetic or radiation) on the x axis and \hop\ column on
the y axis. The contour line corresponds to the observed brightness temperature. This depends on our
assumption for the source size.

\hop\ abundances relative to H$_2$ ($X$(\hop)) are listed in table~\ref{t:column}. We have used CO 2--1 spectra from the literature
and used a Galactic conversion factor from CO luminosity to H$_2$ mass ($2.5 \times 10^{20}$ $\cmmt$ K$^{-1}$ km$^{-1}$ s) to
estimate $N$(H$_2$).  Note that for Galactic nuclei (and starbursts) it is argued that this conversion factor overestimates
the H$_2$ column density by factors of 5-10 (see e.g. \citet{martin10}). Since we wish to reduce the risk of overestimating the \hop\ abundances we
have however still adopted the standard conversion factor. For IC~342 and NGC~253 we have estimated and average $N$(H$_2$) for the 13\arcsec\ beam size.
For NGC~1068, NGC~4418 and NGC~6240 we have used the same adopted source size as for \hop.

\begin{table}
\caption{\label{t:column} Parameters$^a$ for the RADEX Modeling of the \hop\ Line.}
\begin{tabular}{lcccc}
 & \\
\hline
\hline \\

Galaxy & Source Size$^b$ &  $T_{\rm IR}$ & $N$(\hpo)$^c$ fit & $X$(\hop)$^d$ \\
 & [\arcsec ] &  [K] & [$\cmmt$] \\

\hline 
&\\
IC~342 & 13 &  40 & $1.25 \times 10^{14}$ & $4 \times 10^{-9}$ \\
NGC~253 & 13 &  50 & $2.25 \times 10^{15}$ & $1.5 \times 10^{-8}$ \\
NGC~1068 & 2 & 150 & $1 \times 10^{15}$ &  $1 \times 10^{-8}$ \\
NGC~4418 & 0.5 & 80 & $2.5 \times 10^{16}$ & $3 \times 10^{-8}$ \\
NGC~6240 & 1 &  50 & $5 \times 10^{15}$ & $2.5 \times 10^{-8}$  \\	
&\\
\hline 
\\
\end{tabular}

a) The temperature range investigated for all galaxies was 10-500 K, but solutions were fixed where $T_{\rm ex}=T_{\rm IR}$. 
The $N$(\hpo) range searched was $10^{12}-10^{17}$ $\cmmt$.
b) For a discussion of adopted source sizes see section~\ref{s:sample}.
c) The total beam-averaged
column density of \hpo\ is 2-3 times higher than that of p-\hop\ in Fig.~\ref{f:radex}, because the ortho to para ratio of \hpo\ drops from the
high-temperature limit of {\it o/p} = 1 for $T>$ 100 K to {\it o/p}=2 when $T<$ 50 K. Here we have corrected column densities 
for an {\it o/p} ratio of 1.5. 
d) {\it IC~342:} $N$(H$_2$) is estimated to $3 \times 10^{22}$ $\cmmt$ (from the CO data of \citet{eckart90}). 
{\it NGC~253:} $N$(H$_2$) is estimated to $1.3 \times 10^{23}$ $\cmmt$ (using CO data from \citet{mauersberger96}.
{\it NGC~1068:} From \citet{tacconi94} the $N$(H$_2$) for the CND is estimated to be
$2-10 \times 10^{22}$ $\cmmt$. If we take the CO 2--1 data from \citet{planesas89} and assume a source size of 2\arcsec
we obtain $N$(H$_2$)=$1.35 \times 10^{23}$ $\cmmt$.
{\it NGC~4418:} Column density for H$_2$ for a 0.\arcsec 5 CO 2--1 source size is $N$(H$_2$)=$7.4 \times 10^{23}$ $\cmmt$ (Costagliola et al in prep.). 
{\it NGC~6240:} \citet{iono07} estimate $N$(H$_2$) to $1 \times 10^{23}$ $\cmmt$ from their CO 3--2 data.

\end{table}

\subsubsection{Errors in \hop\ column density and abundances.}
\label{s:err}

Our assumptions on the excitation and spatial extent of of \hop\ introduce errors in
the column density estimates. For IC~342 in particular the source size error can be significant.
We assume that the source fills the beam which gives us an average column density for GMCs A, B and E, 
but the real column density towards a particular cloud should be higher. For NGC~253 the double peaked line shape suggests
that the \hop\ emission is emerging from both dense-gas peaks in the center and is filling the JCMT beam.
For NGC~1068 the \hop\ source size in the CND could be anything from $<$1\arcsec\ to 4\arcsec. We have assumed
a source size of 2\arcsec. For NGC~4418 and NGC~6240 the source sizes are $\lapprox$1\arcsec\ which means that
the beam dilution is significant, but no \hop\ emission is missed.

In Fig.~\ref{f:radex}. the impact of the assumption of the excitation temperature on
the resulting \hop\ column density is illustrated. Note the strong dependence of $N$(\hop) for $T_{\rm ex}\lapprox 60$~K.
For NGC~253 (for example) $N$(p-\hop) changes from $9 \times 10^{14}$ $\cmmt$ when $T_{\rm ex}$=50 K to $7 \times 10^{15}$ 
for $T_{\rm ex}$=25 K. When $T_{\rm ex}$ exceeds 60 K, however, the temperature dependence on $N$(\hop) is almost gone.
Thus for all galaxies, except IC~342, we are not likely to overestimate $N$(\hop) by more than a factor of 2 since the
adopted $T_{\rm ex}\gapprox$ 50~K. 
It is possible that \hop\ could be collisionally excited. In this
case the excitation temperature is likely significantly lower than the ones we assume here (beacuse of the high critical density).
In Fig.~\ref{f:radex} it is evident that the $N$(\hop) would go up considerably with decreasing temperature.
With the assumed $T_{\rm ex}$ the resulting $N$(\hop) is already qute large for all galaxies - even larger \hop\ columns would be an interesting result indeed, but very difficult to explain with current models.
The excitation can be constrained through observing multiple transitions and
Requena-Torres et al (in prep.) are currently studying the 307 and 364 GHz line of \hop\ in a sample of starburst galaxies.

We conclude that the source size uncertainties dominate the errors in the column density calculations. 
To improve future $N$(\hop) calculations it is necessary to determine source sizes through high resolution
observations and to obtain information on $T_{\rm ex}$(\hop) through multi-transition observations. 

In addition it should be noted that the estimates of the \hop\ relative abundances are dependent on the reliability of the
conversion factor from CO luminosity to H$_2$ mass.


\section{Discussion}
\label{s:discussion}

\subsection{\hop\ abundances and PDR/XDR models}
\label{s:xdr}

We compare the column densities of \hop\ and H$_2$ in Table~\ref{t:column} to
chemical models of clouds irradiated by either far-UV, photon dominated regions (PDRs)
or X-ray photons (XDRs). In general, we expect
the XDRs to dominate in molecular ISMs surrounding an AGN.
The XDR and PDR models we use are presented in \citet{meijerink05} and \citet{meijerink07}.
In general, the thermal and chemical structure of XDRs and
PDRs are different. Larger parts of an XDR can be maintained at high temperaure
and the ionization fraction of an XDR  can be up to two
orders of magnitude higher ($x_e = 10^{-2} - 10^{-1}$) than in a PDR. In an XDR \hto\ may form in the gas phase
at high temperatures (200-300 K) and then react with either H$_3^+$ or \hcop\ to form \hop.

Source averaged \hcop\ abundances are typically $\approx 10^{-8}$ for all detected galaxies except for
IC~342. A general result of the models presented in \citet{vandertak08} is that the $X$(\hop) does not exceed
$3 \times 10^{-9}$ in PDRs - even in the models with an extremely high cosmic-ray ionization
rate of $\zeta = 5.0 \times 10^{-15}$ $\s-1$. \hop\ abundances approaching $10^{-8}$ and beyond are more
likely to occur in X-ray dominated regions and we conclude that {\it the XDR scenario fits the $X$(\hop)
values of NGC~253, NGC~1068, NGC~4418 and NGC~6240. For IC~342, the
results are consistent with a PDR, but a model with a high $\zeta$ of $5.0 \times 10^{-15}$ $\s-1$.}

\subsubsection{Water abundances}
\label{s:water}
 
The \hop/\hto\ ratio is an even better probe of the ionization rate, and potential nature of the
emitting source. It is straight forward to obtain \hop/\hto\ abundance ratios as large as $10^{-2}$ in XDR models,
while for PDR models ratios are generally $10^{-3}$ or less. {\it Herschel} 
has already measured water abundances in M~82 \citep{weiss10} (see also section~\ref{s:h2op}) and results for other galaxies will follow soon.

\subsection{Relative \hop\ and \hcop\ abundances and XDR/PDR models}
\label{s:hcop}

\citet{kohno01} and \citet{imanishi04} find HCN/\hcop\ 1--0 line intensity ratios greater
than unity in several Seyfert nuclei - where also the HCN/CO 1--0 line ratio is high.  
Furthermore, \citet{gracia06} find elevated HCN/\hcop\ 1--0 line ratios in ULIRGs. This
is often interpreted as a sign of an underabundance of \hcop\ compared to HCN due to
X-ray chemistry. (Although it is important to remember that an abundance difference cannot
be unambiguously deduced from a single-transition line ratio.)
Underabundant \hcop\ in XDRs has been proposed in theoretical work by \citet{maloney96}. However,
more recent models by \citet{meijerink05} suggest that \hcop\ may be underabundant in moderate column
density ($N_{\rm H} < 10^{22.5}$ $\cmmt$) XDRs - but for larger columns the reverse is true
and $X$(\hcop) generally exceeds $X$(HCN).
Here we compare our derived $N$(\hop) with $N$(\hcop) for the sample galaxies to see if large \hop\
abundances are paired with a particularly low \hcop\ abundance, and to compare $N$(\hop)/$N$(\hcop)
with the expectations from current models.

In the XDR/PDR models, the formation of both \hop\ as well as \hcop\
is mainly driven by reactions with H$_2$, H$_2^+$, and H$_3^+$, and destruction
by electrons, unless the electron fractional abundance is very low $x_e \lapprox
10^{-7} - 10^{-8}$. However, the interpretation of an $N$(\hop)/$N$(\hcop) abundance
ratio in an XDR/PDR scenario is not entirely straightforward, but the models provide
some useful limits.
For example, in the PDR models \hop/\hcop\ ratios are unlikely to become
larger than 3 \citep{meijerink10}, while for the XDR models it is quite straightforward to obtain
$N$(\hop)$>N$(\hcop), but the obtained ratio is very column density dependent. When the 
column densities are small (e.g $N$(H) $\approx 10^{22}$ $\cmmt$) ratios as large 20 can be obtained, 
but abundances (and associated brightness temperatures) are small, and in order to obtain significant \hcop\ and 
\hop\ column densities larger clouds are needed. However, when increasing 
the column, the \hop/\hcop\ column density ratio slowly decreases, but 
moderate ratios around 4--5 are easy to reproduce with the XDR model.

\subsubsection{$N$(\hop)/$N$(\hcop) in the sample galaxies}

\begin{table}
\caption{\label{t:hcop} \hcop/\hop Ratios.}
\begin{tabular}{lcc}
 & \\
\hline
\hline \\

Galaxy & $N$(\hcop)$^a$ & $N$(\hop)/$N$(\hcop)\\
       &  [$\cmmt$]     &  \\

\hline 
&\\
IC~342 &  $2-6 \times 10^{14}$ & 0.2 - 0.6 \\
NGC~253 & $5 \times 10^{14}$ & 4 \\
NGC~1068 & $4 \times 10^{13}$  & 24\\
NGC~4418 & $4.7 \times 10^{15}$  & 5\\
NGC~6240 & $3-4 \times 10^{15}$ & 0.7 \\	
&\\
\hline 
\\
\end{tabular}

a) \noindent {\it IC~342:} Estimated from \citet{nguyen92}(Note that a low
\hcop/H$^{13}$CO$^+$ 1--0 line ratio of 7 suggests that \hcop\ columns may be higher.
{\it NGC~253:} From \citet{martin06} ($6.5 \times 10^{13}$ for $T_{\rm ex}$=12 K and for 35\arcsec\ source size.
Rescaled $N$(\hcop) for the 10\arcsec\  by 14\arcsec\  \hop\ source size of
\citep{knudsen07}). 
{\it NGC~1068:} We used the data provided in \citet{jp09} and made a rotational
diagram (assuming a 2\arcsec\ source size) finding $T_{\rm ex}$(\hcop)=36 K.
{\it NGC~4418:} Column density for \hcop\ (based on data in \citet{aalto07b} and unpublished JCMT \hcop\ 4--3 data)
for a 0.\arcsec 5 source size. 
{\it NGC~6240:} We estimate the \hcop\ column density from the
excitation information given in \citet{iono07} and for a source size of 1\arcsec.

\end{table}

In table~\ref{t:hcop} we list $N$(\hcop) and $N$(\hop)/$N$(\hcop) for the sample galaxies.
The largest $N$(\hop)/$N$(\hcop) value is found in NGC~1068 which has a Seyfert nucleus and its inner few hundred pc has
been suggested to be an XDR \citep[e.g.][]{usero04,jp09,garcia10}. The large abundance ratio of $\approx$24 suggested for
NGC~1068 is consistent with (a low column density) XDR.

For NGC~4418 and NGC~253 XDR models can explain the \hop/\hcop\ ratios (but we note that the values with errors are also within the
range of PDR models). For NGC~253 there is evidence that both an AGN and a young starburst is present
while for NGC~4418 the nuclear activity is so obscured that the nature of the activity cannot be discerned (see section~\ref{s:sample}). The \hop/\hcop\ ratio for IC~342 is consistent with a PDR model as is the relative \hop\ abundance.

For NGC~6240 the \hop/\hcop\ ratio is consistent with both XDR and PDR models while the 
\hop\ relative abundance favours an XDR.  It is interesting to note that
even if the molecular gas of NGC~6240 is collected in-between the two nuclei, their radiation could still impact
the gas. The two AGNs are separated by 1\arcsec .5 which means that the molecular gas is irradiated by
intense X-ray emission from two directions on only 370~pc distance. This scenario could possibly explain
both the relatively high \hop\ and \hcop\ abundances. 

\subsection{Grain chemistry and the formation of \hop.}
\label{s:grain}

Above we have found that the XDR/PDR models can potentially explain the \hop\ and \hcop\ column densities
and abundances we found for the observed galaxies. However, for NGC~4418 the problem with an XDR explanation
is the large column densities of gas and dust observed towards its center (see footnote of Tab.~\ref{t:column})
where \citet{costagliola10} find global \hc3n\ abundances
similar to those found towards hot cores in Sgr~B2. Intense vibrational line emission suggest that the \hc3n\
indeed exists in warm-to-hot environments in the very center of the galaxy. The \hc3n\ abundances are not consistent
with either XDRs or PDRs. 

Thus, we have searched for alternative explanations to the \hop\ line emission outside of the XDR and PDR
models. Elevated \hc3n\ line emission is consistent with the conditions in warm, dense shielded gas associated
with embedded star formation.  Is it then possible to obtain relative \hop\ abundances of $\approx 10^{-8}$
without the X-ray chemistry?

In a scenario where grains are important, the \hto\ can evaporate off the grains at temperatures $>$100 K and 
in \citet{phillips92} the destruction of \hcop\ through the reaction of \hto\ is found to be important formation process
for \hop\ when the gas is dense and warm.
For a water abundance $X$(\hto) of $10^{-5}$ and a density $n$(H$_2$) of
$10^5$ $\cmmd$ \citet{phillips92} find an \hop\ abundance, $X$(\hop), of $\approx 10^{-8}$ and $X$(\hcop) a factor 2--3 lower.
It is unclear however, if the $N$(\hop)/$N$(\hto) can be greater than $10^{-3}$ within the context of their model.
\citet{vandertak00} find the same effect in their study of the of Galactic protostar GL~2136.
In their Fig.~1 the temperature and density structure  is presented including calculated
concentrations of \hcop\ both with and without destruction by \hto. 

{\it We suggest that for NGC~4418 it is possible that the \hop\ abundance and the $N$(\hop)/$N$(\hcop) ratio can be explained by a model where the gas is warm and dense and water is coming off the grains to react with \hcop\ to form \hop.} This
process could contribute to a reduction of the \hcop\ abundance in the gas phase which may result in a lower \hcop\ line
intensity in any transition. 
Note that shocks may be responsible for removing the water from the grains making it available for further reactions including
\hop\ formation. \citet{flower10} presents models where water abundances are related to the type and strength of shocks and there are a multitide of Galactic observational studies showing enhancement of water emission in shocks \citep[]{nisini10,lefloch10,wamplfler10,melnick08}. Apart from causing \hto\ to come off the grains, shocks can result in efficient gas-phase formation of \hto\ from OH in the shocked high-temperature regime.

\subsection{H$_2$O$^+$, \hop, and \hto}
\label{s:h2op}

\citet{weiss10} find a (potentially) surprisingly low $N$(\hto)/$N$(H$_2$O$^+$) ratio of only a few which they attribute to \hto\ evaporating off the grains through shocks, the \hto\ then becomes photodissociated into O and OH.
Ion-molecule reactions then make H$_2$O$^+$ while \hto\ abundances remain low. 
It is furthermore noteworthy that $X$(H$_2$O$^+) >$ $X$(\hop) in M~82 \citep{vandertak08,weiss10}.
This is consistent with the low density ($n=10^3$ $\cmmd$), high cosmic ray rate ($> 5 \times 10^{-15}$ s$^{-1}$) model (Model 2) of \citet{meijerink10}. This model does however not include any grain processing. 

In \citet{tak10} a photoevaporation-ionization hypothesis for the \hop/H$_2$O$^+$/\hto\ ratio is discussed for M~82 noting that testing this theory requires calculation of the  photodissociation cross-section of \hop. Whether this hypothesis holds for other galaxies, however, remains to be seen: the \hto/H$_2$O$^+$ ratio in M~82 may well be unusually low. In the ULIRG Mrk~231 
$N$(\hto)$>>N$(H$_2$O$^+$) and the H$_2$O$^+$ line emission feature is consistent with an XDR interpretation \citep{werf10}.

\subsection{Future studies}
\label{s:future}

With this study we have confirmed that the 364~GHz line of p-\hop\ is feasible to detect in
starburst and active galaxies and that \hop\ fractional abundances are substantial.
The HIFI heterodyne spectrometer onboard ESA's {\it Herschel} space observatory is currently
obtaining \hto\ and \hpo\ data around 1~THz and more accurate estimates of
the \hto/\hpo\ abundance ratios can be determined for many sources improving the
understanding of the ionization rates of nuclear molecular
regions. Determining the cource size and excitation of \hop\ is essential to further improve the
understanding of the impact of the nuclear activity on its surrounding interstellar medium and to
facilitate a more accurate comparison with models.
In the near future, grain chemistry and impact of shocks should be added to the interpretation of the 
data. In particular the results for NGC~4418 - and to some degree also NGC~253 - emphasizes this conclusion.



\begin{acknowledgements}
      We are very grateful to the staff of the JCMT for their help and support and we thank the
referee, Santiago Garcia-Burillo, for a thorough and very useful report which improved the paper.
This research has made use of the NASA/IPAC Extragalactic Database (NED) which is operated
by the Jet Propulsion Laboratory, California Institute of Technology, under contract
with the National Aeronautics and Space Administration.
\end{acknowledgements}

\bibliographystyle{aa}
\bibliography{15878aalto}

\end{document}